\documentclass[useAMS,usenatbib,twocolumn]{mnras}

\usepackage[T1]{fontenc}

\DeclareRobustCommand{\VAN}[3]{#2}
\let\VANthebibliography\thebibliography
\def\thebibliography{\DeclareRobustCommand{\VAN}[3]{##3}\VANthebibliography}

\usepackage{graphics}
\usepackage{graphicx}
\usepackage{epstopdf}
\usepackage{amsmath}
\usepackage{amssymb}
\usepackage{xspace}
\usepackage{commath}
\usepackage{times}
\usepackage{soul}
\usepackage{bm}
\usepackage{mathtools}
\usepackage{caption}
\usepackage{graphicx}
\usepackage{verbatim}
\usepackage{changepage}
\usepackage{setspace}
\usepackage{natbib}
\usepackage{braket}
\usepackage{graphicx}
\usepackage{amsmath}
\usepackage{cancel}
\usepackage{caption}
\usepackage{wasysym}

\usepackage{algorithm}
\usepackage{algorithmic}
\usepackage{etoolbox}
\usepackage{cleveref}
\usepackage{physics}
\usepackage{mwe}
\usepackage{wasysym}
\usepackage{hyperref}
\usepackage{cuted}
\usepackage{siunitx}
\usepackage{xspace}
\usepackage{multicol}

\usepackage{acronym}
\newacro{PDF}{probability distribution function}
\newacroplural{PDFs}{probability distribution functions}

\newacro{DF}{distribution function}
\newacroplural{DFs}{distribution functions}

\newacro{BH}{black hole}
\newacroplural{BHs}{black holes}

\newacro{IMBH}{intermediary mass black hole}
\newacroplural{IMBHs}{intermediary mass black holes}

\newacro{VRR}{vector resonant relaxation}

\newacro{SRR}{scalar resonant relaxation}

\newacro{NR}{non-resonant relaxation}

\newcommand{\disperse}{DisPerse\xspace}
\newcommand{\ie}{\emph{i.e.}\xspace}
\usepackage{soul}
\usepackage{xcolor}
\definecolor{bubbles}{rgb}{0.91, 1.0, 1.0}
\definecolor{aquamarine}{rgb}{0.5, 1.0, 0.83}
\definecolor{bubblegum}{rgb}{0.99, 0.76, 0.8}
\definecolor{bluebell}{rgb}{0.74, 0.74, 0.92}
\definecolor{dollarbill}{rgb}{0.72, 0.93, 0.6}
\definecolor{gold}{rgb}{0.99, 0.84, 0.}

\crefname{figure}{Fig.}{Figs}
\Crefname{figure}{Fig.}{Figs}
\crefname{table}{Table}{Tables}
\Crefname{table}{Table}{Tables}
\crefname{equation}{equation}{equations}
\Crefname{equation}{Equation}{Equations}
\crefname{section}{Section}{Sections}
\Crefname{section}{Section}{Sections}
\crefname{appendix}{Appendix}{Appendices}
\Crefname{appendix}{Appendix}{Appendices}

\renewcommand{\vec}[1]{\vb*{#1}}

\newcommand{\rhom}{\bar{\rho}_\mathrm{m}}
\newcommand{\ce}{\mathrm{ce}}
\newcommand{\pk}{\mathrm{pk}}

\newcommand{\heaviside}{\Theta_\mathrm{H}}
\newcommand{\dirac}{\delta_\mathrm{D}}
\newcommand{\orcid}[1]{\href{https://orcid.org/#1}{\includegraphics[height=.7em]{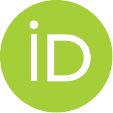}}}

\begin{document}

\title[Analytical model of merger rates and spin]{
Estimating major merger rates  and spin parameters ab initio \\
 via the clustering of  critical events }

\author[ C. Cadiou, E. Pichon-Pharabod, C. Pichon and D. Pogosyan]{
    \orcid{0000-0003-2285-0332} Corentin Cadiou$^{1,2}$\thanks{E-mail: corentin.cadiou@fysik.lu.se},
    Eric Pichon-Pharabod$^{3}$,
    \orcid{0000-0003-0695-6735} Christophe Pichon$^{2,3,4}$
    and
    \orcid{0000-0002-7998-6823} Dmitri Pogosyan$^{5}$\\
\vspace*{6pt}\\
$^1$ Lund Observatory, Division of Astrophysics, Department of Physics, Lund University, Box 43, SE-221 00 Lund, Sweden\\
$^{2}$ CNRS and Sorbonne Université, UMR 7095, Institut d'Astrophysique de Paris, 98 bis Boulevard Arago, F-75014 Paris, France\\
$^3$ IPHT, DRF-INP, UMR 3680, CEA, Orme des Merisiers Bat 774, 91191 Gif-sur-Yvette, France\\
$^{4}$ Korea Institute of Advanced Studies (KIAS) 85 Hoegiro, Dongdaemun-gu, Seoul, 02455, Republic of Korea\\
$^5$ Department of Physics, University of Alberta, 11322-89 Avenue, Edmonton, Alberta, T6G 2G7, Canada
}

\label{firstpage}
\pagerange{\pageref{firstpage}--\pageref{lastpage}}
\maketitle

\begin{abstract}
    We build a model to predict from first principles the properties of major mergers.
    We predict these from the coalescence of peaks and saddle points in the vicinity of a given larger peak, as one increases the smoothing scale in the initial linear density field as a proxy for cosmic time.
   To refine  our results, we also ensure, using a suite of $\sim 400$ power-law Gaussian random fields smoothed at $\sim 30$ different scales,
    that the relevant peaks and saddles are topologically connected: they should belong to a persistent pair before coalescence.     
    Our model allows us to 
    (a) compute the probability distribution function of the satellite-merger separation in Lagrangian space: they peak at three times the smoothing scale;
    (b)  predict  the distribution of the number of mergers as a function of peak rarity: haloes typically undergo two major mergers ($>$1:10) per decade of mass growth;
    (c)  recover that the typical spin brought by mergers: it is of the order of a few tens of percent.
\end{abstract}


\section{Introduction}
\label{sec:intro}

On large scales, the galaxy distribution adopts a network-like structure, composed of walls, filaments, 
and superclusters \citep{Gelleretal89,10.48550/arxiv.2210.16499}. This network is inherently tied to the cosmic microwave background, the relic of the density distribution
in the primordial Universe. The non-uniformity of this initially quasi-Gaussian field evolved under the influence of
gravity into the so-called cosmic web \citep{bkp96} we now observe. One
can therefore hope to predict the evolution of the cosmic web by studying the topological properties of the initial density field. From its evolution, one should be able to predict the rate of mergers
of dark haloes and their geometry hence their contribution to halo spin.

The classical method to study mergers is to run cosmological simulations \citep[e.g.][]{1998ARA&A..36..599B,2020NatRP...2...42V}, compute where haloes are located at each time increment and construct that way their merger tree
 \cite[e.g.][]{1993MNRAS.262..627L,mosteretal13}.

The theory of merger trees for dark haloes has a long-standing history starting from the original Press-Schechter theory
\citep{press_formation_1974}, excursion set \citep{BBKS,PH1990,Bondetal1991} and peak patch theory \citep{Bond1996} or related formalisms \citep{Manrique1995,Manrique1996,hanami,2002MNRAS.331..587M,2022MNRAS.509.5305S}.
One notable recent variation is the suggestion to use peaks of `energy' field as such progenitors \citep{Musso2019}.
One of the prevailing theoretical ideas is to consider peaks of the initial density field in initial Lagrangian space, smoothed at scales related to halo masses, to be halo progenitors \citep{BBKS}.
The statistical properties of mergers can then be predicted analytically through extensions of the excursion set theories \citep{1993MNRAS.262..627L,neisteinMergerRatesDark2008}, or, alternatively, it can be measured in peak-patch simulations \citep{2019MNRAS.483.2236S}.
In the first approach, all the information is local, preventing us from computing the geometry of mergers.
In the second approach, the geometry of mergers is accessible as a Monte Carlo average.

In this paper, we provide an alternative framework that specifically takes into account the geometry of mergers while remaining as analytical as possible.
The goal is not for accuracy, but rather to provide a simple framework to access information such as the statistics of the number of mergers or the spin brought by mergers.

The paper is organised as follows: in \cref{sec:method}, we present our model together with analytical estimates of the number of direct mergers with
a given halo
In \Cref{sec:numerical}, we extend our model to take into account the topology of the density field using persistence pairing via \disperse.
 This allows us to draw mock merger trees and to predict the PDF of the spin parameter that each merger contributes as a function of rarity.
\Cref{sec:conclusion} wraps up.

\Cref{sec:spectral-parameters} introduces the relevant spectral parameters.
 \Cref{sec:first-principle} revisits event statistics while relying on the  clustering properties of peaks and events computed from first principles.
\Cref{sec:event-distrib} recalls the critical event PDF and provides a fit to it. \Cref{sec:mean_evolution} discusses the evolution of  peak rarity with Gaussian smoothing.

\begin{figure}
    \centering
       \includegraphics[width=1.\columnwidth]{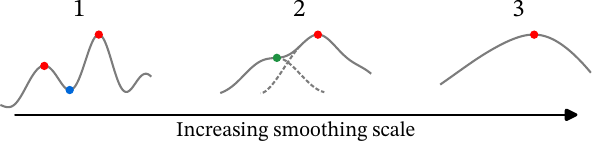}
       \caption{Typical density field around a critical event in 1D. From left to right: 1. We see two distinct objects (in red), separated by a minimum (in blue) in the density field. 2. The critical event occurs (in green), but the density profile still shows two different objects (highlighted with dashed lines). 3. The merger has completed and there remains only one object (in red).
       }
    \label{fig:merger_crit}
\end{figure}
\begin{figure}
\centering
   \includegraphics[width=\columnwidth]{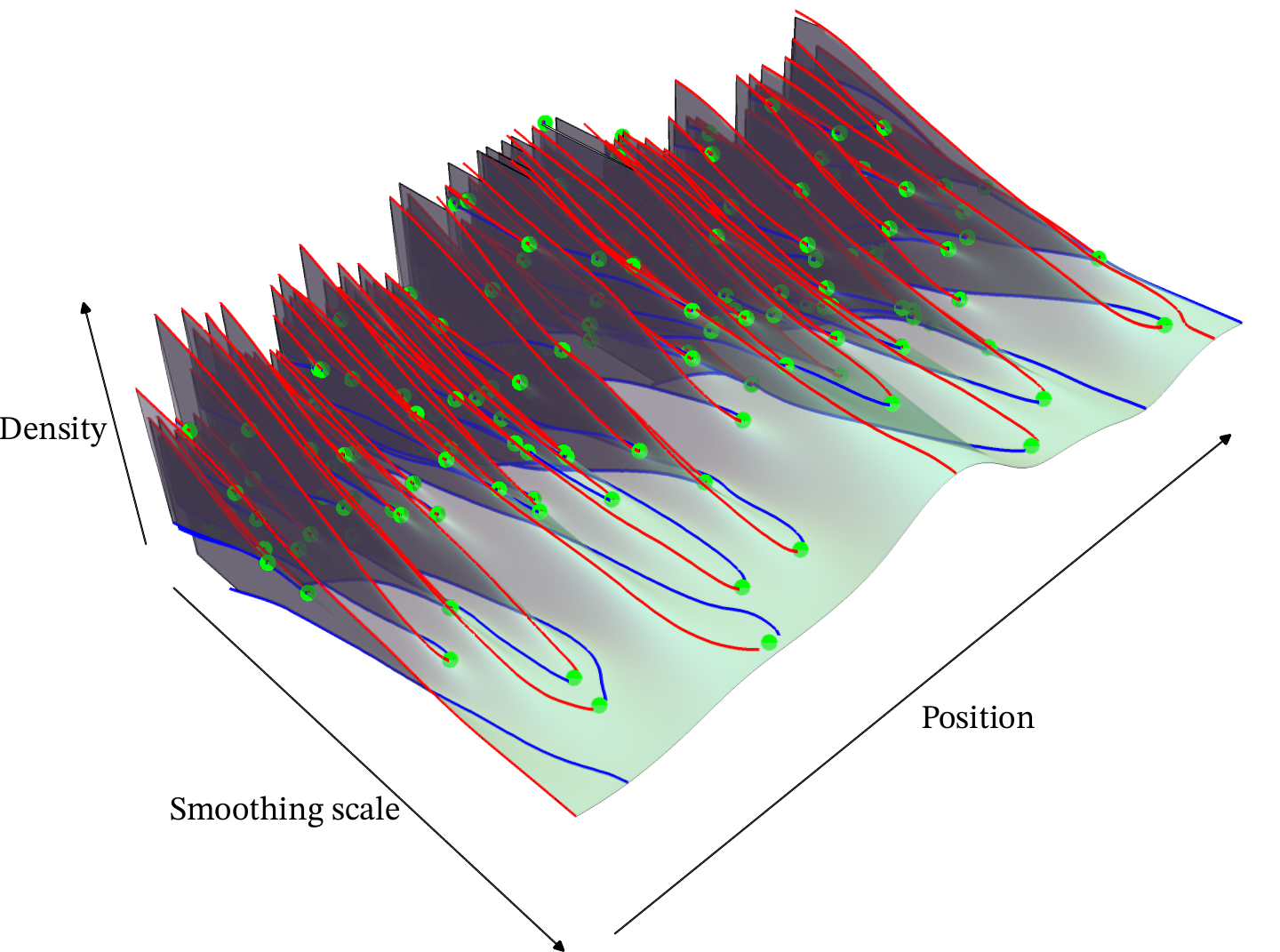}
   \caption{Representation of the smoothing of a 1D Gaussian random field representing density fluctuations in the primordial universe.
   Red lines are maxima of the field, blue lines are the minima. Green points are critical events, where a minimum/maximum pair meets and annihilates itself.
   By counting how many critical events there are in the vicinity of a peak line, one can count the number of subhaloes merging into the corresponding halo.
   }
   \label{fig:1D-smoothing}
\end{figure}
\begin{figure}
    \centering
       \center\includegraphics[width=0.95\columnwidth]{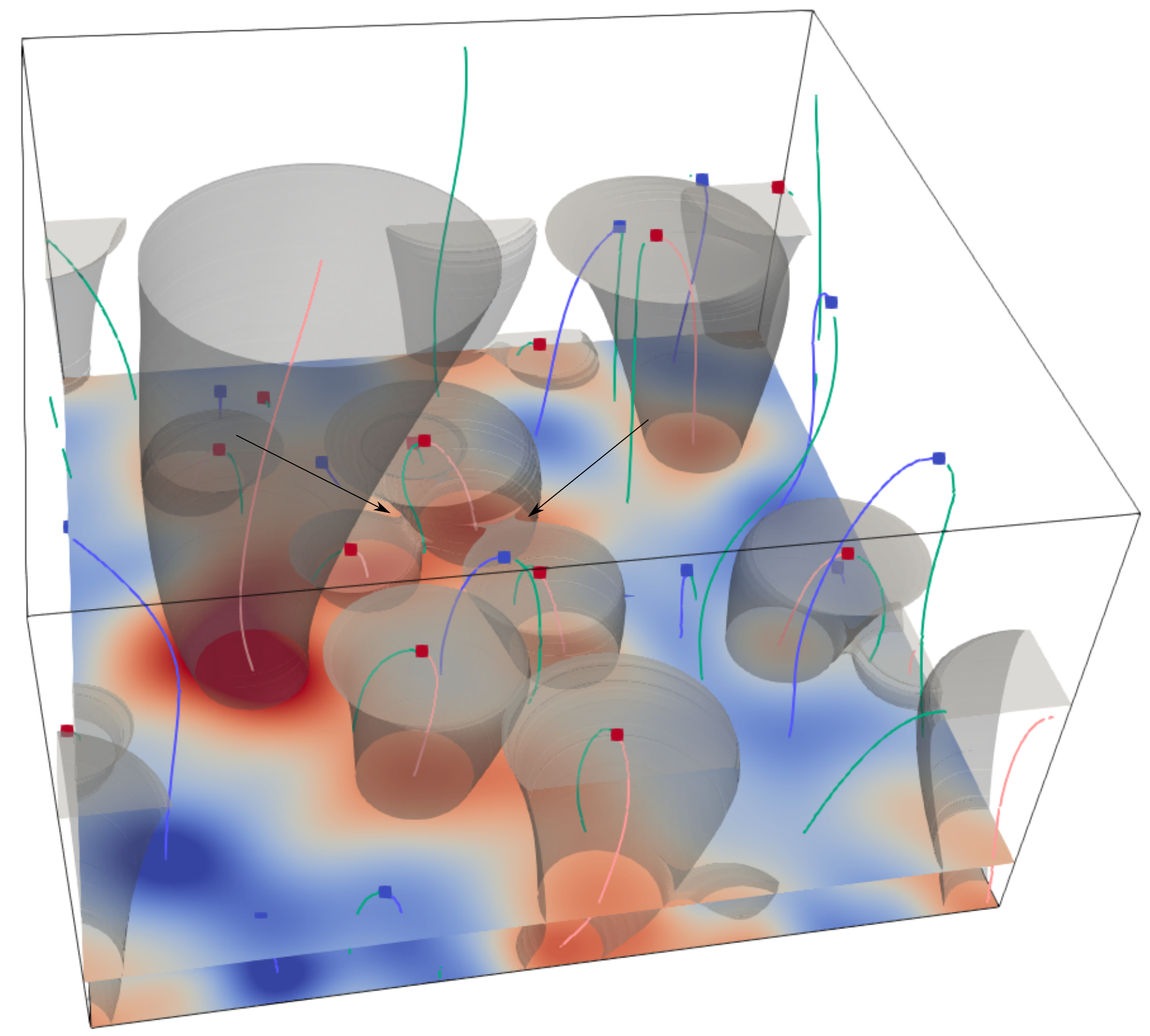}
       \caption{Visualisation of the action of gaussian smoothing on the critical events of a 2D field. The vertical axis is the smoothing scale, increasing upwards.  The horizontal cross-section is a 2D space.
       The various vertical lines represent the tracks of extrema positions (maxima in pink, minima in blue, saddles in green) as one changes the field smoothing.
       The red and blue squares represent the corresponding critical events (in red are points of  peak--saddle and in blue that of minima-saddle coalescence).
        The grey cones show the volume within some fraction of the smoothing scale (here chosen arbitrarily to be 1.2 times the smoothing scale) around each maxima track, which contains all the past physical history of a given peak.
       This paper aims to characterise these cones and the properties of the critical events within them so as to compute major merger rates as a function of final halo mass.
       }
       \label{fig:2D-smoothing}
    \end{figure}

\begin{figure*}
    \centering
    \includegraphics[width=\textwidth]{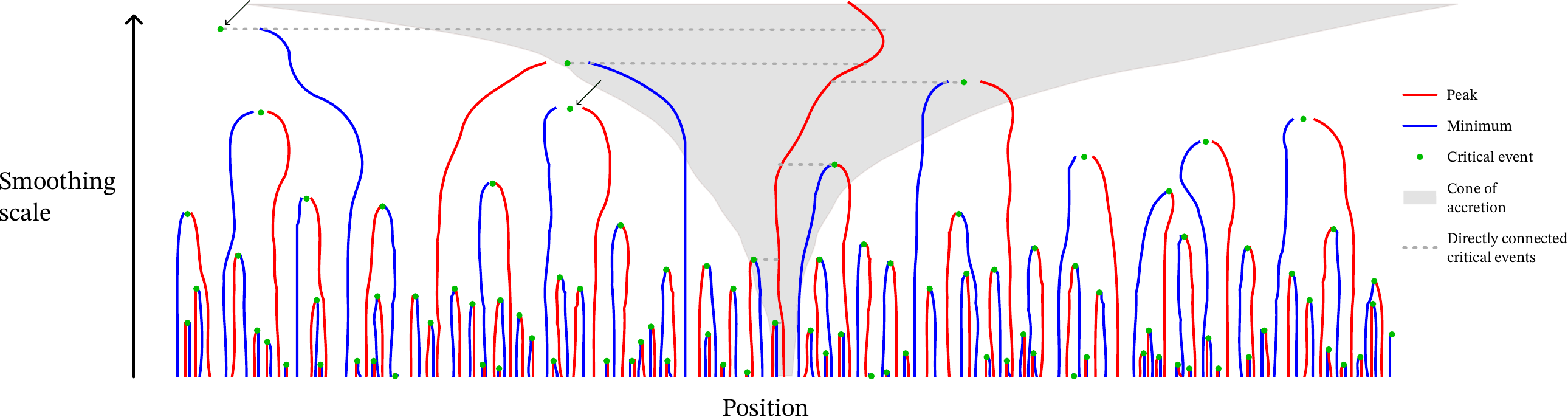}
   \caption{
    Our model associates peaks (in red) to haloes, and critical events (in green) to mergers as a function of smoothing scale $R$ which encodes mass evolution.
    Critical events correspond to the coalescence of a peak with a minimum (in blue).
    We count any critical event directly connected to the central peak (\ie{}~not separated by another minimum) as a merger.
    Here, this definition would give  6 mergers (shown 
    with grey dashed lines) into the central peak.
Alternatively, we can estimate this number by counting the number of critical events within the shaded grey ``influence cone'' of spatial width $\alpha R$.
Here, $\alpha=3.3$ is chosen so that the cone includes only directly connected events capturing five events.
Note that the left-most event indicated by an arrow is missed,
but it lies at large smoothing $R$ where the sample variance is large. 
The second arrow gives an example of an event that is not directly connected to the central
peak, instead merging with the peak on its left. The cone boundary separates this event from 
the directly connected one above it.
}\label{fig:crit_paths_tracking_cone}
\end{figure*}

\section{Analytical model of mergers per halo}\label{sec:method}

\subsection{Model of halo mergers: cone of influence of a halo}
\label{sec:cone-of-influence}

The starting point of our approach is the peak picture  \citep{BBKS}.
In this framework, peaks in the linear density field are the seeds for the formation of haloes.
We rely on the proposition that the overdensity $\delta(R)$ of a given peak smoothed at scale $R$ can be mapped to a formation redshift $z$, using the spherical collapse model, $\delta(R_ \mathrm{TH}) D(z) = \delta_{\mathrm{c}}$, where $ \delta_{\mathrm{c}}=1.686$ is the critical density, $D(z)$ is the growing mode of perturbations and $R_\mathrm{TH}$ is the top-hat scale.  The mass of the halo formed at this time is inferred from the top-hat window scale, $M = \rhom \frac{4}{3} \pi R_\mathrm{TH}^3$.

While the peak-patch theory provides useful insight into the origin of haloes of a given mass at a given time, it becomes singular during mergers. Indeed, by construction, the merging halo disappears into the larger one together with its corresponding peak.
Here, we instead rely on the analysis of the geometry of the initial linear density field in $(N+1)$D, where $N$ is the dimension of the initial Lagrangian space and the extra dimension is the smoothing scale.
At each smoothing scale, we can formally define critical points (peaks, saddles, and minima) following~\cite{BBKS}.
As smoothing increases, some of these critical points eventually disappear when their corresponding halo (for peaks) merge together.  This approach allows to capture mergers in space-time from an analysis of the initial conditions in position-smoothing scale employing the critical event theory \citep{hanami,Cadiou2020},
where critical events of coalescence between peaks and saddle points serve as proxies for merger events.
The process is sketched in \cref{fig:merger_crit} and illustrated in $(1+1)$D in \cref{fig:1D-smoothing}.
In the following, we will rely on the use of Gaussian filters, which allow
us to advance further the analytical description.
First, this choice allows us to obtain analytical results with the critical event theory.
Second, it yields peaks whose density decreases monotonically with smoothing scale ---~and hence
 the collapse time of the associated halos grows under the spherical collapse model.
While exploring the effect of different filters is beyond the scope of this
paper, we note that using any filter that is positive with
sufficiently smoothly tapered boundaries does not pose any fundamental challenge to the numerical analysis we perform later on. Tophat filter by itself
is not of this class, having sharp boundaries leading to ill-defined second- and
third-derivatives of the smoothed field with cold dark matter-like 
power spectrum.%

Let us track the Lagrangian history with decreasing smoothing of one peak first identified at a smoothing scale $R_0$ and position $x_\pk(R_0) $.
Since we employ Gaussian filters, we need to match the Gaussian and Top-Hat smoothing scales, $R$ and $R_{\mathrm{TH}}$ to assign mass to haloes. The criteria of equal mass encompassed by the filter%
\footnote{Different criteria modify the relation between Gaussian and Top-Hat filters, for instance
	matching the variance of the perturbations leads to $R_{\mathrm{TH}} \approx 2.1 R$. Here, we instead rely on matching masses 
within the filtering 
	window $W$, which
implies $\int W_\mathrm{TH}(r,R_\mathrm{TH}) \dd[N]{r} = \int W_\mathrm{G}(r,R_\mathrm{G}) \dd[N]{r} $.}
gives $R_\mathrm{TH} \approx 1.56 R$ in 3D.
At $R_0$ the peak describes a halo that has collected its mass from a spherical cross-section of $(N+1)$D space of volume $\propto R_0^N$; we can call this sphere a Lagrangian patch of the halo.
At smaller $R$, the Lagrangian position of a peak changed to $x_\pk(R)$ and the volume of its Lagrangian patch decreased
to $\propto R^N$.  The history of the peak in the $(N+1)$D space, including mass accumulation,   now consists of its trajectory
$x_\pk(R)$ and a cone of cross-section volume $\sim R^N$ around it
as shown in \cref{fig:2D-smoothing} for $(2+1)$D example.

A critical event marks the end of a trajectory of a peak that disappeared when its scale reached $R_\ce$ and is absorbed into a surviving peak.
Counting all critical events within a $(N+1)$D straight cylinder with a spatial cross-section of radius $R_0$ around the final surviving peak position $x_\pk(R_0)$ will give the number of all mergers that ever happened within this peak Lagrangian patch. For instance, if two small haloes have merged together before merging with a larger halo, that would count as two events.
However, we are interested in counting only the last \textit{direct} merger event that brought the combined mass of the two small haloes into the main one. Physical intuition tells us that to count only those, we need to
count critical events within the history cone of the peak mass accumulation. Indeed, the Lagrangian patch of the peak grows in size by absorbing the layers along its boundary, and if that layer contains a critical event, it is a direct merger.

In $(1+1)$D, as illustrated in \cref{fig:crit_paths_tracking_cone}, direct mergers correspond to critical events (green points)
that are not separated by any intervening minimum (blue line) from the surviving peak.
In $(N+1)$D, this is generalized by only counting a critical event as a direct merger if, at fixed smoothing $R=R_\ce$, it is connected 
to the main peak in $N$D space by a filamentary ridge  with no other saddles in-between.
\cref{fig:crit_paths_tracking_cone} confirms that, indeed, 
one can find a ``cone of influence'' around a peak line 
which spatial cross-section radius is $\alpha R$, where $\alpha$ is between
2 and 4, that contains most and, vice versa,
almost exclusively the critical events
that are directly connected to this peak. 
The very fact that such a cone can be defined, at a statistical
precision, even in principle, is non-trivial.  The reason for that is,
as \cref{fig:crit_paths_tracking_cone}
demonstrates again, the presence of the exclusion zone around the main peak
devoid of any other critical events. This causes direct merger events to be most likely located in 
the first layer of critical events from the main peak. 
This exclusion effect is expected from peak-peak correlation studies \cite[e.g.][]{2021PhRvD.103h3530B}
and can be used as a quantitative way to determine the boundary 
of the ``cone of influence'' 
that we develop in  \cref{sec:clustering} and \cref{sec:first-principle}.

Counting critical events within the ``influence cone'' of the peak that
 contribute to the mass and spin growth of this
surviving halo is the main analytical tool of this paper.%

The evolution of a halo in smoothing direction tells us directly its history in terms of mass accumulation, as $M \propto R^{N}$. That is, we can describe what happened with the halo as its mass increased, say, 10-fold.   In the next
section, we apply this picture to count merger events.
We shall limit ourselves to the $(3+1)$D case as it bears the most relevance to dark matter halo formation. We however note that there is no additional theoretical difficulty in deriving the general $(N+1)$D case.

\subsection{Number of major merger events within a mass range}
\label{sec:merger-count}

Here and in the following, we are interested in counting the number of objects that directly merged into an object of final mass $M_0$ as its mass grew from $f^3 M_0$ to $M_0$, where $f < 1$.
Since smoothing scale maps directly onto mass, this amounts to counting the number of critical events between two scales $R_0 \propto \sqrt[3]{M_0}$ and $R_1 \propto f \sqrt[3]{M_0}$.

Let us now count direct \textit{major} mergers that we define as mergers with satellites that bring at least $f^3 M_0$ mass in the merger event.
First, note that we define here the mass ratio with respect to the final mass of the peak at a fixed time, rather than at the time of the merger.
The number of such mergers, as halo grew from $f^3 M_0$ to $M_0$,
is given by the number of critical events in the section of the
cone of influence of the halo contained between  $f R_0$ and $R_0$. 
It can be obtained with the following integral
\begin{equation}
    \label{eq:merger_rate_gen}
    N_\mathrm{merger} = \int_{f R_0}^{R_0} \dd{R} \int_{0}^{\alpha R} \dd[3]{r} n_\ce(R,\vb*{r}),
\end{equation}
where $n_{\ce}(R,\vb*{r})$ is the number density of critical events at the point $(R,\vb*{r})$ in the extended $(3+1)$D space of positions--smoothing scale 
and the adjustable parameter $\alpha$, introduced in the previous section,
sets the radial extent of the $R=\mathrm{const}$ cross-section of the influence cone.
Note that in the above formula, we are able to avoid the dependence on the past trajectory $x_\pk(R)$ of the halo, by treating each slab of constant $R$ independently, and by evaluating the radial distance $r$ to the critical event found of at $R$ from the main peak position, $x_\pk(R)$, defined at the same smoothing.

As a first estimate, let us approximate $N_\mathrm{merger}$ by taking the density of the critical events inside the cone of influence to be equal to its mean value.
Thus, we first neglect any spatial correlation between the existence of a peak (corresponding to the surviving object in a merger) and the critical event (corresponding to the absorbed object in a merger).
In~\cite{Cadiou2020}, we determined that the average density of critical events in $(3+1)$D that correspond to peak mergers is given by
\begin{equation}
\label{eq:mean_nce}
\bar{n}_{\ce} = \frac{R}{R_*^5} \frac{1-\tilde{\gamma}^2}{\tilde{\gamma}^2} \frac{29\sqrt{15}-18\sqrt{10}}{600 \pi^2}\,,
\end{equation}
where the spectral scale $R_*$ and parameter $\tilde{\gamma}$ are given in \cref{sec:spectral-parameters}, together with other parameters that characterize the statistics of a density field.
Note that $R_* \propto R$, so the number density of critical events scales as $n_{\ce} \propto R^{-4}$. The cubic part of this dependence,  $R^{-3}$, reflects the decrease of spatial density of critical points with increasing smoothing scale; the additional scaling, $R^{-1}$, reflects that the frequency of critical events is uniform in $\log R$.

Assuming a power-law density spectrum with index $n_\mathrm{s}$ and Gaussian filter,  \cref{eq:mean_nce} gives
\begin{equation}
\bar{n}_{\ce} \approx 0.0094 \left(\frac{5+n_{\mathrm{s}}}{2}\right)^{3/2} R^{-4}\,.
\end{equation}
Using this mean value $\bar{n}_{\ce}$ for critical density in \cref{eq:merger_rate_gen}  gives us a first rough, but telling estimate of the
merger number that a typical halo experiences
\begin{equation}
\label{eq:N_m_crude}
N_\mathrm{merger} \approx 0.039 (-\ln f) \left(\frac{5+n_{\mathrm{s}}}{2}\right)^{3/2}  \alpha^3.
\end{equation}
The final results depend on the specific values of $f$ and $\alpha$.
The value of the fraction $f$ controls down to which mass ratio mergers should be included and has thus a straightforward physical interpretation.
The parameter $\alpha$ controls the opening of the cone of influence around the peak from which critical events should be considered as direct mergers and is to be determined.

Using \cref{eq:N_m_crude} with $n_\mathrm{s}= -2$, which corresponds to the typical slope of a $\Lambda$CDM power-spectrum at scales ranging from Milky-Way-like systems to clusters, we can count the number of direct mergers with mass ratio larger than 1:10 ($f=1/\sqrt[3]{10}$).
For the choice of $\alpha$, a natural idea would be to
 count only
critical events within the halo Top-Hat Lagrangian radius, then
$\alpha=R_\mathrm{TH}/R\approx 1.56$ and we find 
$N_\mathrm{merger} \approx 0.2$.
However, this is clearly an underestimation for $\alpha$, 
as it corresponds to the situation where the center of the merging halo has already been
incorporated into the main one when the latter was of equal mass.
A more sensible choice, as will bear out in the analysis presented in the
next section is to extend the opening of the cone to twice that ratio 
$\alpha=2 R_\mathrm{TH}/R \approx 3.1$ which corresponds to the two mass spheres of
the surviving and merging peaks touching at the scale of the critical event, signifying the merger onset. We  then find $N_\mathrm{merger} \approx 1.7$.
Overall, we see that the number of direct major mergers that a halo experiences while increasing its mass tenfold is small, of the order of one or two.  For scale invariant history, each decade of mass accumulation contributes a similar number of mergers, so a cluster that grew from a galactic-scale
protocluster thousand-fold in mass ($f=1/\sqrt[3]{1000}$), did it having experienced $N_\mathrm{merger} \approx 5$ major mergers.
These conclusions follow directly from first principles while studying the structure of the initial density field.

\subsection{Accounting for rarity and clustering}
\label{sec:rarity-clustering}

\begin{figure}
    \centering
    \includegraphics[width=\columnwidth]{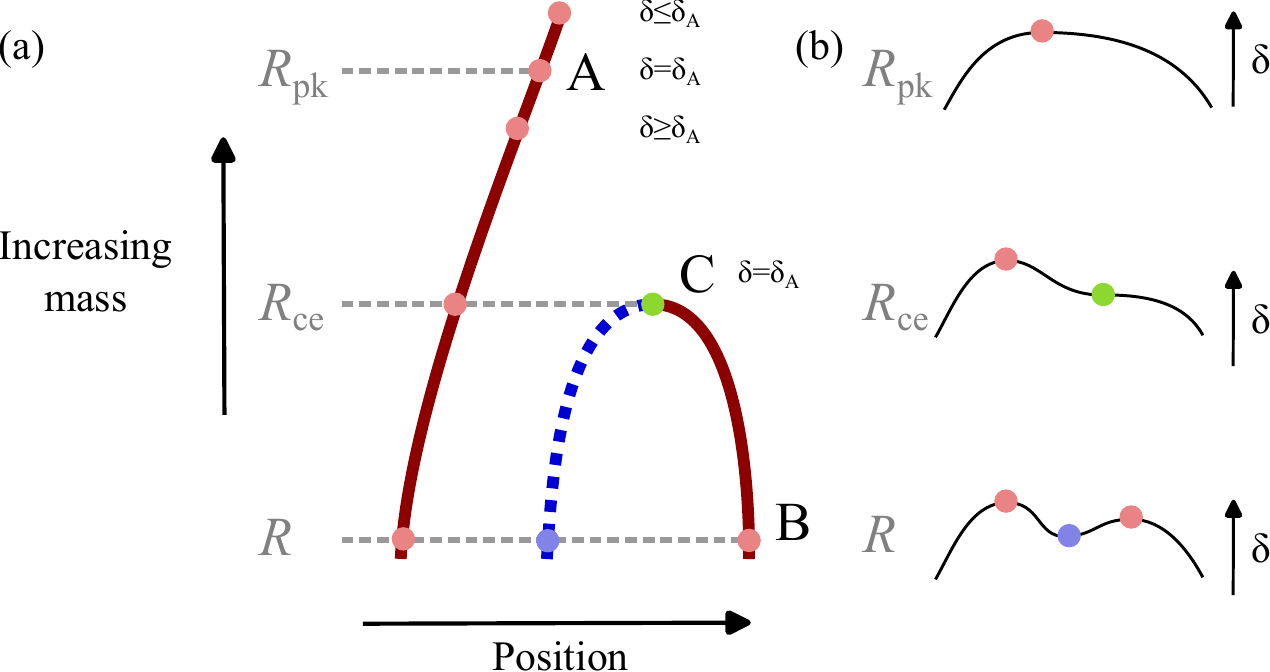}
    \caption{
        Panel a) shows the tracks of two peaks (red) and a saddle point (blue) of the density field as a function of smoothing scale $R$ in 1+1D space.
        Panel b) shows the field profile at three smoothing scales.
        At smoothing scale $R_0$, the peaks and saddle points are distinct. At smoothing $R_\ce$, one peak and a saddle point create a critical event C, after each
       only one peak survives to larger smoothing scales.   The merging of peaks  is completed at the scale $R_\pk$
       when the overdensity of the surviving peak, now at point A,  is equal to the overdensity of the critical event $\delta_\mathrm{C}$ and thus can be viewed as
       reaching the threshold $\delta_\mathrm{c}$ for halo formation at the same time.
        We can then interpret the critical event as a merger with mass ratio $M_\mathrm{A}/M_\mathrm{C}=(R_\ce/R_\pk)^3$ at a redshift corresponding to $D(z)=\delta_\mathrm{c}/\delta_\mathrm{A}$.
    }\label{fig:model}
\end{figure}

Let us now refine our model so as to define $\alpha$ more rigorously by i) requiring the merging object to have gravitationally collapsed before it merges and by ii) taking into account the correlations between the central halo peak and the merger critical events.

Let us therefore consider the number density of critical events of given height $\nu$ as a function of their distance $r$ to a central peak, both defined at the same smoothing scale $R$
\begin{equation}
n_{\ce|\pk} (\nu,R,r) =  \bar{n}_\ce(R) C(\nu,\gamma)  \left(1+\xi_{\ce,\pk} (\nu,R,r) \right)\,,
\label{eq:condne}
\end{equation}
where $C$ is the distribution function, $\int_{-\infty}^\infty C(\nu,\gamma) \dd{\nu}=1$, of critical events in overdensity and is given by the analytical formula in \cref{sec:diff_events} along with a useful Gaussian approximation.  The clustering of critical events in the peak neighbourhood is described by the peak-critical event correlation function on a slice of fixed $R$, $\xi_{\ce,\pk}(\nu,R,r) $.  The composite index $\pk$ refers to any peak parameters that may be specified as a condition. Our goal is to determine what range of $\nu$ and
what extent of $r$ one needs to consider to count haloes merging into a surviving halo with a particular $\nu_\pk$.

\subsubsection{Heights of critical events}
\label{sec:heights}

 To establish the range of critical event heights that describe mergers of real haloes, we rely on the spherical collapse approximation to map the peak overdensity to its gravitational collapse time. 
 We consider a physical halo at redshift $z$ to be described by a peak at the Top-Hat scale $R_\pk$ such that $\delta_\pk(R_\pk) = \delta_{\rm c}/D(z)$, where $\delta_{\rm c}$ is the critical overdensity for collapse model and $D(z)$ is the linear growing mode value at redshift $z$. 
 Thus, a critical event found at smoothing $R_\ce$ describes the merger of a satellite halo of scale $R_\ce$ into the main peak at scale $R_{\pk}$ that corresponds to the same redshift, i.e such that $\delta_{\ce}(R_\ce)=\delta_\pk(R_\pk) = \delta_{\rm c}/D(z)$. The situation is demonstrated in \cref{fig:model}. The ratio of scales and, correspondingly, masses of the two merging haloes is therefore determined by the condition of equal overdensities at the time of the merger.
 
  Let us now consider a peak that has reached a scale $R_0$ and that experienced a merger in its past when its scale was $R_\pk$, $R_\pk \le R_0$. Requiring that, at the time of the merger, the surviving peak's scale (mass) is larger than that of the satellite sets the relation $ R_0 \ge R_\pk \ge R_\ce$ or, conversely, $ \sigma_0(R_0) \le \sigma_0(R_\pk) \le \sigma_0(R_\ce$), where $\sigma_0$ is the r.m.s. of the density field and is defined in \cref{sec:spectral-parameters}. From which it follows that the rarity of the relevant critical events at scale $R_\ce$, 
 \begin{equation}
	 \nu=\frac{\delta_\ce(R_\ce)}{\sigma_0(R_\ce) }
	 = \frac{\delta_\pk(R_\pk)}{\sigma_0(R_\ce)} = \nu_\pk \frac{\sigma_0(R_\pk)}{\sigma_0(R_\ce)}
	  \end{equation}
 is within the range
  \begin{equation}
	  	\label{eq:vce_range}
	  	 \nu_\pk \frac{\sigma_0(R_0)}{\sigma_0(R_\ce)} \le \nu \le \nu_\pk.
	 \end{equation}
The lower bound corresponds to mergers that have been completed at the very last moment,
$R_\pk = R_0$. The upper bound is achieved for mergers of two equal mass objects, $R_\pk = R_\ce$.

To obtain the total number of merger events in a halo history we now integrate the conditional event density in \cref{eq:condne} over the physically relevant range of heights from \cref{eq:vce_range}, \ie
\begin{multline}
	\label{eq:merger_rate_exact}
	N_\mathrm{merger}(\nu_\pk) = \\
	\int_{f R_0}^{R_0} \dd{R}
	\int_{0}^{\infty} 4\pi r^2 \dd{r}
	\int_{\nu_\pk \frac{\sigma_0(R_0)}{\sigma_0(R)}}^{\nu_\pk}\dd{\nu}
	n_{\ce|\pk}(\nu,  R, r).
\end{multline}

\subsubsection{Clustering of critical events around peaks}
\label{sec:clustering}
Ultimately, the conditional event density $n_{\ce|\pk}$ should include only critical events that will directly merge with the peak; this would make the density of critical events go to zero $n_{\ce|\pk}(\nu,R,r) \to 0 $ far from the peak, \ie{} when $ r \to \infty$.
While we cannot implement this condition analytically (as it is non-local), we can measure $n_{\ce|\pk}(\nu,R,r)$ numerically, as will be done in the upcoming \cref{sec:numerical}.
We can however approximate the conditional density {\sl ab initio} by relaxing the conditions that the peak is linked to the critical event to obtain $\widetilde{n}_{\ce|\pk}$ given by
\begin{equation}\label{eq:peak-event-corellation}
    \widetilde{n}_{\ce|\pk}(\nu, R, r) \equiv 
\frac{\langle \operatorname{Peak}(\vb*{x}) \operatorname{Event}(\vb*{y})\rangle}{\langle \operatorname{Peak}(\vb*{x})\rangle}\,,
\end{equation}
using formally straightforward analytical calculations of  critical event -- peak correlations, 
as described in \cref{sec:first-principle}.
Here, brackets denote an ensemble average, where
$\vb*{x}$ and $\vb*{y}$ are the random vectors containing the density and its successive derivatives at the location of the peak and the critical event, respectively 
and `$\operatorname{Peak}$' and `$\operatorname{Event}$' enforce the peak and critical event conditions respectively.
We can expect $\widetilde{n}_{\ce|\pk}$ 
to track the exact $n_{\ce|\pk}$  up to several smoothing length distances from the peak, but further away from the peak it just describes the mean unconstrained density of critical events: $\widetilde{n}_{\ce|\pk}(\nu,R,r) \to \bar{n}_{\ce}(R) C(\nu)$ as $r \to \infty$.  Therefore,  in this approximation, the question of where to truncate the integration over the peak neighbourhood remains  and we have
\begin{multline}
    \label{eq:merger_rate_tilde}
    N_\mathrm{merger}(\nu_\pk) \approx \\
    \int_{f R_0}^{R_0}\dd{R}
    \int_{0}^{\alpha R} 4\pi r^2 \dd{r}
    \int_{\nu_\pk \frac{\sigma_0(R_0)}{\sigma_0(R)}}^{\nu_\pk}\dd{\nu}
    \widetilde{n}_{\ce|\pk}(\nu, R, r).
\end{multline}

For scale-free spectra, introducing the dimensionless ratios $u=r/R$ and $w=R/R_0$ and changing the order of integration, \cref{eq:merger_rate_exact} can be written as
\begin{align}
    \nonumber
    N_\mathrm{merger}(\nu_\pk) &=
    \int_{f }^{1} \frac{\dd{w}}{w}
    \int_{0}^{\infty} \dd{u} \frac{\dd[2]{N_{\ce|\pk}}}{\dd u \; \dd \ln w}, \\
    \frac{\dd[2]{N_{\ce|\pk}}}{\dd u \; \dd \ln w} &\equiv 4 \pi u^2
    \int_{\nu_\pk w^{(n+3)/2}}^{\nu_\pk}\!\!\!\!\! \dd{\nu} w^4 {n}_{\ce | \pk} (\nu,w,u)\,,
\label{eq:dNdudw}
\end{align}
where we note that the differential event count per logarithm of smoothing scales has only a dependence on the smoothing scale $w$ \emph{via} the bounds of the height integration, since for power-law spectra $w^4 n_{\ce|\pk}(w)$ is $w$-independent.

\begin{figure}
  \includegraphics[width=\columnwidth]{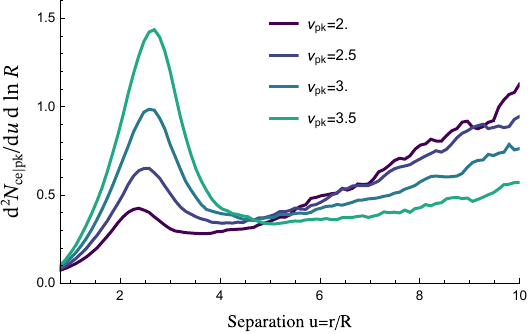}
    \caption{
        Integration of the analytical expression of the clustering of critical events around peaks, \cref{eq:dNdudw}, for different values of $\nu_\pk$. Here $\nu_\ce$ is integrated from $0.681\nu_\pk$ to $\nu_\pk$, with power spectrum $n=-2$ and smoothing ratio $R_\pk/R_\ce = 0.95$.
        We detail the method used to compute this graph in \cref{sec:monte_carlo}.
   Rarer peaks have a stronger excess of events in their vicinity, which mirrors
   the known Kaiser bias \citep{Kaiser1984}.
    }\label{fig:2pcf_analytical}
\end{figure}

In \cref{fig:2pcf_analytical}, we plot the differential version of \cref{eq:dNdudw} per radial spatial shell, integrated
over $\nu$ in the range $0.681 \nu_\pk \le \nu \le \nu_\pk$ which corresponds to $(R/R_0)^3=1/10$
for $n=-2$ power-law spectrum.
\cref{fig:2pcf_analytical} shows that critical events are indeed preferentially clustered at distances $\sim 3 R$ from
a peak.  It is natural to interpret this distance as the boundary of the cone influence of the peak, with more distant critical events constituting ``the field'' of events that are not directly merging into the peak. Thus,
the computation of the correlation function supports choosing $\alpha \approx 3$ in the estimate of \cref{eq:N_m_crude}, \ie{} qualitatively double the value of $R_\mathrm{TH}$.
Note also that the probability of a critical event to be found close to a peak is suppressed, which corresponds to an exclusion from the interior of the peak's cone of influence, as was qualitatively observed in \cref{fig:crit_paths_tracking_cone}.

\section{Numerical integration of mergers} \label{sec:numerical}

We will now rely on a numerical approach to study the properties of the mergers a given halo undergoes through the properties of critical events that are absorbed by the peak representing the halo. The main numerical step is the identification of critical event--peak pairs that mark specific mergers.  This in turn involves the identification in the initial field at a sequence of smoothing scales of connected peak--saddle--peak triplets that describe filamentary links between the merging peaks. At a particular smoothing scale, the saddle and one of the peaks of a given triplet merge into a critical event, while the remaining peak is a surviving halo.  Such triplets are identified with \disperse \citep{sousbie09} that performs Morse complex analysis of the random density field.
We perform the analysis on an ensemble of scale-free Gaussian realizations of initial density fields.

\subsection{Number of mergers per peak line}

We identify proto-events as close-to-zero persistence-pairs of peaks and filament-saddles,
and we study here their clustering relative to a given peak.
We proceed as follows: we generate 430 scale-invariant gaussian random fields of size $256^3$ with a power law power spectrum with spectral index $n=-2$ .
We smooth each field with a Gaussian filter over multiples of a given smoothing scale. Here we choose $R=1.1^k\ \si{pixel}$ with $k$ running from 15 to 49.  Each smoothing step corresponds to \SI{33}{\percent} increase
in the mass associated with the smoothing scale, and 8 steps give the mass increasing approximately 10-fold.
The skeleton is identified at each scale using \disperse \citep{sousbie09} with a persistence threshold of \SI{1}{\percent} of the RMS of the smoothed fields.
We also measure at each critical point the value of the gravitational acceleration.
Since we rely on Gaussian filters, the overwhelming majority of critical events encode the destruction of a peak with a saddle with increasing smoothing, rather than the creation of such a pair (the latter is $30\times$ less likely, see~\citealt{Cadiou2020}, Figure F1).
In the following, we ignore the latter rare case as it has no straightforward physical interpretation and is subdominant.
We thus detect events by following each pair of peak and saddle from one smoothing scale to the next one.
Each pair either survives at the next smoothing scale or disappears at a critical event.
In the latter case, we define the position of the critical event as the middle point of the pair at the largest smoothing scale where it survived.
Given that \disperse{} \citep{Sousbie2011} stores for each saddle the two maxima it is paired with, we can therefore associate the critical event to the one peak that is not involved in the event.

\begin{figure}
    \includegraphics[width=\columnwidth]{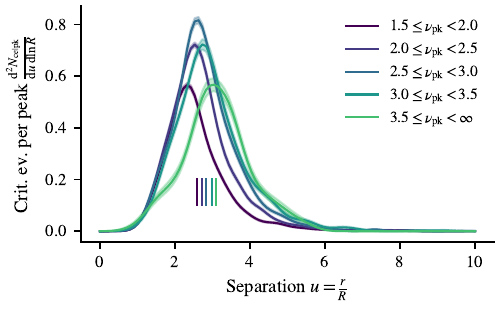}
    \caption{
        The radial number density of topologically linked critical event per peak as a function of distance to the peak on a slice of constant $R$ and per logarithmic bin in smoothing scale
        (see the text for details of how critical events are identified and paired to a peak).
        We represent the mean separation with vertical lines.
        Critical events are preferentially clustered at $r \sim 2.5-3.5 R$, depending on rarity. When compared to \cref{fig:2pcf_analytical}, the rarer peaks have relatively fewer connected events, suggesting that their satellites would have merged together before merging into the central object.
    }\label{fig:Ncepk_Npk}
\end{figure}

Using this data, we can now study the clustering of critical events around peaks.
In \cref{fig:Ncepk_Npk}, we show the ratio of the number of critical events per logarithmic smoothing scale bin per unit distance to the number of peaks at the same scale and at the same rarity\footnote{Note that some peaks will not be associated to any critical event in the fixed interval of smoothing scales.}.
Critical events are counted in the interval of rarity $0.681 \; \nu_\pk \leq \nu_\ce \leq \nu_\pk$. This range corresponds to the smallest scale cross-section of the influence cone
if we count major mergers of a peak that grows ten times in the process,
\ie{}  $R_\ce = 1/\sqrt[3]{10} R_0$ in \cref{eq:merger_rate_exact}. 
 Note that our quantity differs from two-point correlation functions in which one would compute the distance between any critical event and any peak, as was done in the previous section (\cref{fig:2pcf_analytical}).
Here, each critical event only contributes to the density at a single distance to its associated peak, and we thus expect the signal to drop to zero at infinite separation since the probability for a critical event to be associated with a peak far away vanishes.

Should the critical events be randomly distributed, their number per linear $\dd r$ would grow as $r^2$. Instead, in \cref{fig:Ncepk_Npk} we find an exclusion zone at small separations, an excess probability at $r/R \sim 2-3$, and a cut-off at large separations as a consequence of our requirement for the critical event to be paired to a peak.

We integrate the curves in \cref{fig:Ncepk_Npk} and report 
the mean number of critical events per peak per logarithmic smoothing bin in \cref{tab:Ncepk_Npk}.  All the effects accounted for in our numerical analysis give just $\approx \SI{40}{\percent}$ lower
values in \cref{tab:Ncepk_Npk} relative to \cref{eq:N_m_crude} (per $\ln f$) with $\alpha=3.1$.  This shows that the two main corrections to the naive uniform density estimate --- the restriction to only collapsed satellite haloes
(\cref{sec:heights}) on the one hand, and an attraction of critical events of similar heights towards the peak
influence zone (\cref{sec:clustering}) on the other hand --- compensate each other to a large extent.

As a function of peak rarity $\nu_\pk$, the mean number of critical events per peak in a constant $R$ slab first increases up to $\nu_\pk \approx 3$ before decreasing. It is fairly flat in the range $\nu_\pk = 2.5-3.5$, where most of the physical 
haloes are, which argues for taking $\alpha$ parameter independent on $\nu_\pk$ if one uses the rough estimate as the global mean density of events times effective volume as in \cref{eq:N_m_crude}, \cref{sec:merger-count}.  
\begin{table}
	\centering
	\caption{
		Number of critical events per peak, both being measured at the same scale, per logarithmic bin of smoothing scale, as a function of the rarity of the peak.
	}\label{tab:Ncepk_Npk}
	{
		\addtolength{\tabcolsep}{-1.1pt}
		\begin{tabular}{cc cc cc}
			$\nu_\pk$ range & $[1.5,2.0[$ & $[2.0,2.5[$ & $[2.5,3.0[$ & $[3.0,3.5[$ & $[3.5,\infty[$\\[2pt]
			\hline
			\\[-1.5\tabcolsep]
			$\dfrac{\dd{N_{\ce|\pk}}}{\dd{\ln R}}$ & 0.90 & 1.20 & 1.40 & 1.42 & 1.27
		\end{tabular}
	}
\end{table}

So far, we have only studied properties of peaks and critical events at the same scale, using the defining \cref{eq:vce_range} to \emph{implicitly} perform a multi-scale analysis.
We can however track peaks from one slab of smoothing scale to another to build peak lines and associate them to critical events.
This generalizes the procedure sketched on \cref{fig:crit_paths_tracking_cone} in 3+1D.
For each peak with mass $M_\pk$, we follow its peak line to find all associated critical events with a mass $M_\pk \geq M_\ce \geq M_\pk / 10$ that have $\delta_\ce(R_\ce) \geq \delta_\pk(R_\pk)$ where $R_\ce$ and $R_\pk$ are now different.
We only retain peaks that exist at scales $R_\pk \geq R_\mathrm{min} \sqrt[3]{10}$ to ensure that we do not miss critical events below our smallest scale.

\begin{figure}
    \includegraphics[width=\columnwidth]{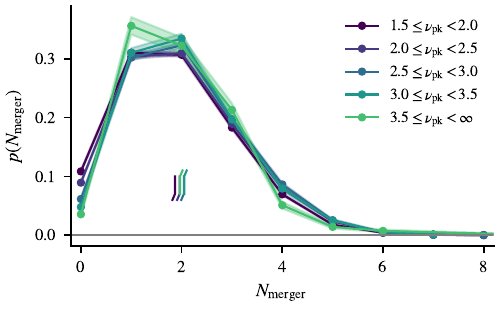}
    \caption{
        Distribution of the number of major mergers for peaks in different rarity bins, as labeled.
        The mean number of major mergers increases from $1.8$ for low-$\nu$ peaks to $2$ for high-$\nu$ peaks and is graphically represented with vertical lines with tilted caps for readability.
    }\label{fig:number_of_mergers}
\end{figure}

Given our numerical sample, we are now in a position to study the distribution of the number of mergers per peak, which we show on \cref{fig:number_of_mergers}.
This measurement differs from the value one would obtain by taking \cref{tab:Ncepk_Npk} multiplied by the logarithm of the range of scales.
Indeed, we account here for the decrease in the number of relevant satellites as their mass approaches the final peak mass.
We also obtain our measurement by computing the number of mergers per surviving peak at the scale $M_\pk$.
Compared to that, in \cref{tab:Ncepk_Npk}, we give the value per any peak at $1/10 M_{\pk}$ scale, irrespective of whether and how long it would survive at the further smoothing scales. 
We find that the dependence of the number of mergers with peak height is weak, with a mean number of mergers varying between 1.8 and 2.  This shows that in the language of effective influence cone radius of \cref{eq:N_m_crude},
$\alpha=2 R_\mathrm{TH}/R=3.1$ is a very good proposition. 

\subsection{Orbital spin parameter of mergers} \label{sec:pdf-spin}
\begin{figure}
    \includegraphics[width=\columnwidth]{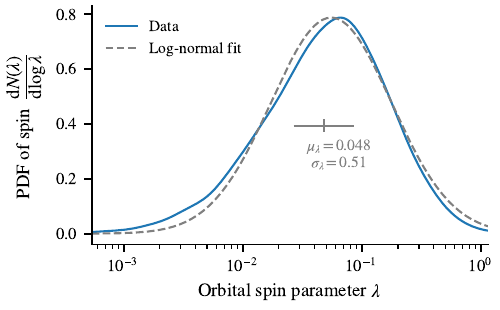}
    \caption{
        Merging objects bring in orbital spin; we quantify here its distribution.
        The distribution resembles a log-normal distribution with parameters $\mu_\lambda \approx 0.048$ and $\sigma_\lambda \approx 0.51$; the distribution resembles its $N$-body counterpart.
    }\label{fig:spin}
\end{figure}

Let us now define the spin parameter of an event at scale $R_\ce$ relative to a peak of rarity $\nu$ at scale $R_\pk$ as
\begin{equation}
    \lambda = \frac{R_\ce^3}{R_\pk^3} \frac{\left| \vb*{r} \crossproduct (\grad \psi - \grad\psi_\mathrm{pk}) \right|}{\sqrt{2} R_\pk \sigma_{-1}(R_\pk)}.
    \label{eq:lambda_def}
\end{equation}
Here, we recall that $\sigma_{-1}$ is the variance of the gradient of the potential, whose definition is given in \cref{sec:spectral-parameters}.
$\grad\psi$ and $\grad\psi_\mathrm{pk}$ are the gravitational accelerations at the locus of the critical event and of the peak respectively, and $\vb*{r}$ is the position of the critical event with respect to the peak.
This definition reflects the spin definition of \cite{bullock_UniversalAngularMomentum_2001a}:
under the Zel'dovich approximation, $R_\ce^3|\vb*{r}\crossproduct(\grad \psi - \grad\psi_\pk)|$ is proportional to the mass times the cross product of the position with the velocity of the merging object relative to the peak at the time of its collapse, \ie{} the numerator is proportional to the orbital angular momentum of the merger.
Conversely, the denominator is proportional to the orbital angular momentum a merger of mass $\propto R_\pk^3$ coming from a distance of $R_\pk$ with velocity equal to the r.m.s. of the velocity $\sqrt{2}\sigma_{-1}$ would have at the time of the collapse of the peak.%
We show on \cref{fig:spin} the distribution of the spin parameter $\lambda$.
The sample consists of mergers with mass ratios greater or equal to 1:10 around rare peaks.
In practice, we select $\nu_\pk > 2.5; R_\pk/\sqrt[3]{10}\leq R_\ce\leq R_\pk$ and $\delta_\ce \geq \delta_\pk$.
The distribution is roughly log-normal with a mean value of $\mu_\lambda \approx 0.048$ and a standard deviation of $\sigma_\lambda \approx 0.51$.

    The novelty of our approach is to model mergers as punctual events.
    This has to be compared to previous theoretical works which treated angular momentum accretion around peaks as a continuous accretion process \mbox{\citep[e.g.][]{white84,rydenGalaxyFormationRole1988}}.
    This allows direct comparisons to results about mergers obtained from $N$-body simulations.%
Remarkably, our estimate for the orbital spin of mergers is found to be very close to the values measured in $N$-body simulations,
which are on the order of 0.04 \citep[see e.g.][]{bullock_UniversalAngularMomentum_2001a,aubertetal04,danovich_FourPhasesAngularmomentum_2015}.
    This translates two effects. First, the mass ratio ($R_\ce^3/R_\pk^3$) is dominated by the more numerous smaller mergers, which drives $\lambda$ down. In addition, we find that critical events tend to have a mostly radial acceleration so that their tangential velocity is smaller than the r.m.s. of the velocity field ($(\grad\psi-\grad\psi_\pk)/\sigma_{-1}$). This is qualitatively in line with findings from $N$-body simulations \citep{wetzelOrbitsInfallingSatellite2011,jiangOrbitalParametersInfalling2015}, where the tangential velocity of major mergers was found to be smaller than their radial velocity.%
Albeit simplistic, our model thus allows us to provide a natural explanation for the fact that mergers bring in a comparable amount of angular momentum to that of the full halo:
gravitational tides alone (which is the only ingredient of our model) can funnel in a significant amount of angular momentum through mergers.
This provides a theoretical motivation for the amplitude of the spin jumps during mergers employed in semi-analytical models \citep[see e.g.{}][]{Vitvitska2002,benson_RandomwalkModelDark_2020}.

\subsection{Mass distribution of mergers}\label{sec:mass}

\begin{figure}
    \includegraphics[width=\columnwidth]{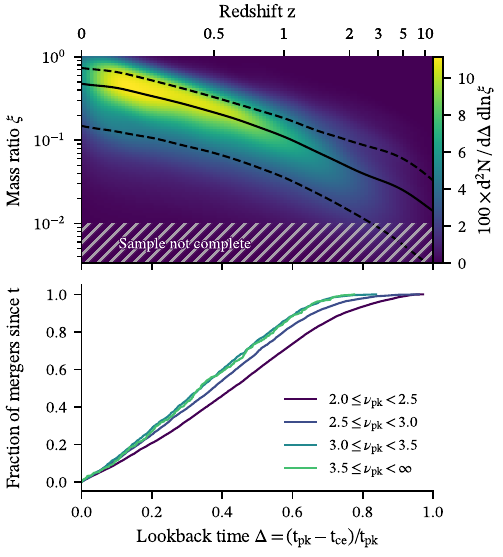}
     \caption{
        Top: distribution of the merger as a function of mass ratio and lookback time.
        We show the median with \SI{68}{\percent} interval in black.
        The hashed area corresponds to regions of the parameter space that may not be complete, see the text for details.
        Bottom: the corresponding fraction of major merger as a function of lookback time relative to the peak collapse time, for different peak heights.
        Rarer objects have had more recent mergers.
     }\label{fig:mass}
\end{figure}

Finally, we compute the mass distribution of mergers as well as their density.
The goal here is to obtain the distribution of the time and mass ratios of the mergers.
In order to build the distribution of the mass ratio, we track peaks over multiple decades of smoothing scales (\ie{} mass).
Since the number density of peaks evolves as $R_\pk^{-3}$, the sample size quickly decreases with smoothing scale; we select here peaks that exist at a scale larger than $R_\mathrm{min} \sqrt[3]{100}\approx 4.6 R_\mathrm{min}$.
The practical consequence is that our sample is only complete over two decades in mass ratio.
We also only retain rare peaks that have $\nu_\pk >2.5$.
Let us then estimate the time of the merger as follows:
we associate both the peak and the critical event a time $t_\pk$ and $t_\ce$ respectively, using $\delta = \delta_\mathrm{c} / D(z)$.
Note that some of the selected peaks will have collapsed by $z=0$ while others will in the distant future.
To aggregate peaks collapsing at different times, we compute the lookback time of the merger relative to the collapse time of the peak $\Delta = (t_\pk - t_\ce) / t_\pk$.
We estimate the distribution of mergers in lookback time-mass ratio space using a 2D kernel density estimation which we show in \cref{fig:mass}, top panel.
As expected, the more massive the merger, the more recently it happened.
    In the figure, we included mergers with mass ratios smaller than $0.01$ in the shaded region rather than truncating the plot, but we remind the reader that our dataset is not complete in this region.%

We then show on the bottom panel of \cref{fig:mass} the cumulative distribution function of the merger time, for mass ratios larger than 1:10 as a function of peak height.
We recover the trend found in $N$-body simulations that rarer haloes have had more recent major mergers than lower mass ones.
Cluster-like structures ($\nu_\pk\gtrapprox 3.5$) typically had \SI{80}{\percent} of their mergers in the second half of their life (past \SI{7}{Gyr} for a cluster at $z=0$), and had half their mergers in the last third of their life (last \SI{5}{Gyr} for a cluster at $z=0$).

Our model reproduces qualitative trends observed in $N$-body simulations \citep[figure 9]{genel_halo_2009,fakhouri_merger_2010}, namely that rarer peaks typically had their last major merger more recently than less rare ones.

    Our model differs from the Extended Press-Schechter (EPS) approach \citep{Bondetal1991,bower1991,1993MNRAS.262..627L}.
    In the EPS approach, random walks in $\sigma(R)-\delta(R)$ are mapped onto mass and time evolution respectively.
    Jumps from one $R$ to another at fixed overdensity thus correspond to sudden changes in mass at a fixed time, \ie{} mergers.
    While this approach yields statistically useful merger rates, it is unable to provide spatial information about mergers since neither the satellite nor the central halo are localized.
    Our model trades some of the analytical tractability of the EPS approach for additional information about the spatial location of mergers, all the while yielding a good agreement with $N$-body simulations.%

\section{Conclusion and perspectives}
\label{sec:conclusion}

We built a model of mergers from an analysis of the initial conditions of the Universe.
Following the work of~\cite{hanami,Cadiou2020}, we relied on the clustering of critical events around peaks in the initial density field to study the properties of halo mergers.
We started with a simple model that yields analytically tractable results, while further refinements presented in \cref{sec:numerical} allowed for more precise results at the cost of numerical integration.

We focussed here on the analysis of merger events that bring at least \SI{10}{\percent} of the \emph{final} mass of the main halo, which we refer to as `major mergers'.
We however note that our approach can be extended to the analysis of minor mergers, which we leave for future work.

We first obtained a zero$^\text{th}$-order analytical estimate of the mean number of mergers per decade in mass using the mean abundance of critical event per peak in \cref{sec:cone-of-influence,sec:merger-count}.
Our results are consistent with haloes having had one to two major mergers {per relative decade of mass growth}.
We then refined our model in \cref{sec:rarity-clustering}  by accounting for the timing of the collapse of the haloes involved in a merger candidate; a critical event should only be counted as a merger if its two associated haloes have collapsed before the merger happens.
We showed this can be achieved semi-analytically by numerically computing the value of a cross-correlation function, \cref{eq:merger_rate_tilde}, that reveals that critical events cluster at 2-3 times the smoothing scale of the peak (\cref{fig:2pcf_analytical}).

Finally, in \cref{sec:numerical}, we addressed the double-counting issue, whereupon a given critical event may be associated with several peaks, by uniquely associating each to the one peak it is topologically connected to.
To that end, we relied on multi-scale analysis of Gaussian random fields using computational topology to restrict ourselves to the study of peaks that, up to the critical event, form persistent pairs.
In this model, we found again that haloes of different rarities undergo about 2 major mergers.
By tracking peaks in position-smoothing scale and by associating critical events with them, we were able to provide numerically cheap and easy-to-interpret data on the statistical properties of halo mergers. We found that mergers come from further away for rarer peaks  (\cref{fig:Ncepk_Npk}), but that the total number of major mergers only weakly depends on peak rarity
(\cref{fig:number_of_mergers}).

We then computed the gravitational tides at the location of the critical event to estimate the relative velocity of mergers and predict the orbital spin they bring in. 
We find that it has a log-normal distribution with a mean of $\mu_\lambda = 0.048$ and $\sigma_\lambda = 0.5$.
These properties are remarkably close to the distribution of DM halo spins measured in hydrodynamical simulations \citep[$\mu_\lambda = 0.038$, $\sigma_\lambda = 0.5$,][]{danovich_FourPhasesAngularmomentum_2015}.
This suggests that our model captures the (statistical properties) of the orbital parameters of mergers, as is expected should they be driven by gravitational tides \citep{cadiouAngularMomentumEvolution2021,cadiouStellarAngularMomentum2022}.
We also computed the distribution of the mass brought by mergers and their timing, which we found to be in qualitative agreement with results obtained in $N$-body simulations \citep{fakhouri_merger_2010}.

While the aim of this model was not to compete with numerical simulations, it provides theoretical grounds to explain the properties of mergers observed in $N$-body simulations and efficient tools to predict their statistics and geometry \emph{ab initio}.
Our model could be improved with precision in mind, for example by taking into account deviations from spherical collapse under the effect of shears to improve our time assignments.
It however reveals that the statistical properties of the merger tree of dark matter halo can be explained through a multi-scale analysis of these initial conditions. 

\subsection{Perspectives}

 Statistics involving successive mergers could potentially be built on top of our model by using critical events associated with the same peak line, for example, to study the relative orientation of the orbital angular momentum of successive mergers.
However, we found that such analysis was complicated by the fact that peaks move with smoothing scale. Different definitions of angular momentum (distance to the peak at the same scale, at the same density, or for a fixed peak density) yielded qualitatively different results. This should be explored in future work.

The model built in this paper relied on a linear multi-scale analysis of the density field.
This could be employed to provide control over the merger tree (timing of the merger, orientation) in numerical simulations through `genetic modifications' of the initial field \citep{roth_GeneticallyModifiedHaloes_2016,rey_QuadraticGeneticModifications_2018,stopyraGenetICNewInitial2020,cadiouAngularMomentumEvolution2021,cadiouCausalEffectEnvironment2021}.
We also note that, as we did for tidal torque theory in \cite{cadiouStellarAngularMomentum2022}, such an approach would allow direct testing of the range of validity of the model.

    In our analysis throughout the paper, we used
    Gaussian filters
    which provide closed analytical formulas for the critical event theory
    and also ensure that, along the peak trajectory, larger smoothing scales 
    correspond to later redshifts of the collapse. 
    In addition, Gaussian filters limit the extent to which peak-saddle point pairs can be created (rather than destroyed, see \citealt{Cadiou2020}), for which we have no physical interpretation yet.
    However, other choices of filter could be considered, as long as the filter
    is positive, so that the region associated with the peak at a larger smoothing
    incorporates the regions assigned with smaller smoothing, and has 
    sufficiently smoothly tempered boundaries to define second and
    third derivatives of the field. 
    For instance, the widely used Top-Hat filter has a compact support and 
    allows to directly map the density
    to a collapse time and the smoothing scale to a mass through the spherical
    collapse model. But it has sharp boundaries and ill-defined derivatives
    beyond
    the first one for a density field with typical cosmological 
    power spectra in CDM-like hierarchical models. For critical event theory, 
    one could consider a modification of Top-Hat that retains a compact
    support with nearly equal 
    weight but has at least second derivatives continuous at the boundary. 
    Closed analytical formulas for the critical event theory
    with such non-Gaussian filters are not known, however, we do not see
    any obstacle to numerical analysis using them.
    The focus of this work was to provide a theoretical framework to understand the properties of mergers, and we expect the qualitative results to hold for other reasonable filters.%

This paper focused on mergers of peaks corresponding to the relative clustering of peak-saddle events.
One could extend the analysis to the relative clustering of saddle-saddle events to provide a theoretical explanation for which filaments merge with which, thus impacting their connectivity or their length \citep{galarragaEspinosaEvolutionCosmicFilaments}.
 Conversely, extending the model to the relative clustering of
 saddle-void events (which wall disappears when?) is also of interest, 
 as the latter may impact spin flip, and is dual to void mergers, and as such could act as a cosmic probe for dark energy.
One could compute the conditional merger rate subject to a larger-scale saddle-point as a proxy to the influence of the larger-scale cosmic web, following both \cite{Musso2018} and \cite{Cadiou2020} to shed light on how the cosmic web drives galaxy assembly \citep{kraljic_galaxies_2018,laigle_COSMOS2015PhotometricRedshifts_2018,hasanEvolvingEffectCosmic2023}.
Eventually, such a theory could contribute to predicting the expected rate of starburst
or AGN activity as a function of redshift and location in the cosmic web. 
\section*{Acknowledgements}

We thank  S.~Codis and S.~Prunet for their early contributions to this work via the co-supervision of EPP's master, and J.~Devriendt and
M.~Musso for insightful conversations.
We also thank the KITP, which was supported in part by grant NSF PHY-1748958, for hosting the workshop \href{https://www.cosmicweb23.org}{`\emph{CosmicWeb23: connecting Galaxies to Cosmology at High and Low Redshift}'} during which this project was advanced.
This work is partially supported by the grant
\href{https://www.secular-evolution.org}{Segal ANR-19-CE31-0017}
of the French Agence Nationale de la Recherche and by the National Science Foundation under Grant No. NSF PHY-1748958.
CC acknowledges support from the Knut and Alice Wallenberg Foundation and the Swedish Research Council (grant 2019-04659).
We thank Stéphane Rouberol for running smoothly the
infinity Cluster, where the simulations were performed.

\section*{Author contributions}
The main role of the authors was, using the CRediT (Contribution
Roles Taxonomy) system (\href{https://authorservices.wiley.com/author-resources/Journal-Authors/open-access/credit.html}{https://authorservices.wiley.com/author-resources/Journal-Authors/open-access/credit.html}):

{\bf CC}: Conceptualization; formal analysis; investigation; methodology; software; writing -- Original Draft Preparation; supervision.
{\bf EPP}: Investigation; software; writing -- Review \& Editing; Visualisation.
{\bf CP}: Conceptualization;  formal analysis; methodology; software; investigation; writing -- Review \& Editing; supervision.
{\bf DP}: Conceptualization; formal analysis; methodology; writing -- Review \& Editing; validation; supervision.

\section*{Data availability}
The data underlying this article will be shared on reasonable request to the corresponding author.

\bibliographystyle{mnras}
\setlength{\bibsep}{-1.0pt} %
\renewcommand*{\bibfont}{\normalsize} %

\bibliography{main}

\appendix

\section{Notations} \label{sec:spectral-parameters}

Let us first introduce the dimensionless quantities for the density field, smoother over a scale $R$ by a filter function $W$
\begin{equation}
    \delta(\vec{r},R) = \int\frac{\dd[3]{k}}{(2\pi)^3} \delta(\vec{k})W(kR)e^{i\vec{k}\cdot\vec{r}}.
\end{equation}
We will consider the statistics of this field and its derivatives in this paper. For practical purposes, let us introduce
\begin{equation}
    \sigma_i^2(R) \equiv \frac{1}{2\pi^2}\int_0^\infty \dd{k}k^2P(k)k^{2i}W^2(kR). \label{eq:defRMS}
\end{equation}
These are the variance of the density field $\langle \delta^2\rangle = \sigma_0^2$, of its first derivative $\langle \grad{\delta}\cdot \grad{\delta}\rangle = \sigma_1^2$, etc.

Following \cite{pogo09b}, let us introduce the characteristic scales of the field
\begin{equation}
    R_0 = \frac{\sigma_0}{\sigma_1}, \quad R_* = \frac{\sigma_1}{\sigma_2}, \quad \tilde{R} = \frac{\sigma_2}{\sigma_3}. \label{eq:defR0}
\end{equation}
These scales are ordered as $R_0 \geq R_* \geq \tilde{R}$. These are the typical separation between zero-crossing of the field, mean distance between extrema, and mean distance between inflection points \citep{BBKS,Cadiou2020}. Let us further define a set of spectral parameters that depend on the shape of the underlying power spectrum. Out of the three scales introduced above, two dimensionless ratios may be constructed that are intrinsic parameters of the theory
\begin{equation}
    \gamma \equiv \frac{R_*}{R_0} = \frac{\sigma_1^2}{\sigma_0\sigma_2}, \quad \tilde{\gamma} = \frac{\tilde{R}}{R_*} = \frac{\sigma_2^2}{\sigma_1\sigma_3}.
    \label{eq:defgamma}
\end{equation}
From a geometrical point of view, $\gamma$ specifies how frequently one encounters a maximum between two zero-crossings of the field, while $\tilde\gamma$ describes, on average, how many inflection points there are between two extrema. These scales and scale ratios fully specify the correlations between the field and its derivative (up to third order) at the same point. For power-law power spectra of index $n$, $P(k)\propto k^n$, with Gaussian smoothing at the scale $R$ in 3D, we have $R_0 = R\sqrt{2/(n+3)}$, $R_* = R\sqrt{2/(n+5)}$ and $\tilde{R} = R\sqrt{2/(n+7)}$ while $\gamma = \sqrt{(n+3)/(n+5)}$ and $\tilde{\gamma} = \sqrt{(n+5)/(n+7)}$.

\section{Peak-event correlation }\label{sec:first-principle}

In order to compute the number of mergers in the vicinity of a peak, we need to compute the two-point correlation function between critical events and peaks.
We achieve this by evaluating the joint density of peaks and critical events
\begin{equation}
    n_\mathrm{peak,ce}=
    \left\langle\operatorname{Peak}(\vb*{x}) \operatorname{Event}(\vb*{y})
    \right\rangle,
\end{equation}
where the average is taken over all 30 random variables defining the peak and critical event field up to the third derivative.
\begin{align}\label{eq:peak-event-char}
    \nonumber
    \operatorname{Peak}(\vb*{x}) \equiv& |J_x| \heaviside(-\mathrm{Tr}(\vb{H_x}))\heaviside({\textstyle\sum_i} \vb H_{x,i})\heaviside(\det\vb H_x)\\
    &\times \dirac(x_1)\dirac(x_2)\dirac(x_3)\dirac(x-\nu_\pk),\\
    \nonumber
    \operatorname{Event}(\vb*{y}) \equiv& |J_y| \heaviside(-\mathrm{Tr}(\vb{H_y}))\heaviside({\textstyle\sum_i} \vb H_{y,i})\dirac(\det \vb H_y)\\
    &\times \dirac(y_1)\dirac(y_2)\dirac(y_3) \dirac(y-\nu)\,.
\end{align}
Here, $\vb*{x}=\{x,x_1,\dots,x_3,\vb{H}_x,\vb{H}_{x,1},\dots,\vb{H}_{x,3}\}$ with $x$, $x_i$, $\vb{H}_x$, $\vb{H}_{x,i}$ the density, its gradient, its Hessian and the minors of its Hessian at the peak location and $\vb*{y}=\{y,y_1,\dots,y_3,\vb{H}_y,\vb{H}_{y,1},\dots,\vb{H}_{y,3}\}$ with $y$, $y_i$, $\vb{H}_y$ and $\vb{H}_{y,i}$ the density, its gradient, its Hessian and the minors of the Hessian at the critical event location. $|J_x|=|\det \vb{H}_{x}|$ and we provide an explicit formula for $|J_y|$ in \cref{eq:J_critevent}. This expression is the rotationally invariant equivalent of Equation 18 of \cite{Cadiou2020}.

 The two-point correlation function can then be found as
\begin{equation}
	\label{eq:xi_pk_ce}
	1 + \xi_{\ce,\pk}(\nu_\pk,\nu,R,r) = \frac{ \left\langle\operatorname{Peak}(\vb*{x}) \operatorname{Event}(\vb*{y})
		\right\rangle}{\langle \operatorname{Peak}(\vb*{x}) \rangle \langle \operatorname{Event}(\vb*{y}) \rangle}.
\end{equation}

Note that, compared to \cite{Cadiou2020}, we cannot perform the integration in the frame of the Hessian here. Indeed, the numerator involves cross correlation between the peak and the critical event which breaks the rotational invariance assumption.
While the exact integration cannot be carried out analytically, we can nonetheless compute it numerically.

\section{Covariance matrices}
\label{sec:covariance_matrices}
The covariance matrix of $x$, $x_i$, $x_{ij}$ is given by
\begin{equation}
    C=
\begin{pmatrix}
 1 & 0 & 0 & 0 & \frac{\gamma }{3} & \frac{\gamma }{3} & \frac{\gamma }{3} & 0 & 0 & 0 \\
 0 & \frac{1}{3} & 0 & 0 & 0 & 0 & 0 & 0 & 0 & 0 \\
 0 & 0 & \frac{1}{3} & 0 & 0 & 0 & 0 & 0 & 0 & 0 \\
 0 & 0 & 0 & \frac{1}{3} & 0 & 0 & 0 & 0 & 0 & 0 \\
 \frac{\gamma }{3} & 0 & 0 & 0 & \frac{1}{5} & \frac{1}{15} & \frac{1}{15} & 0 & 0 & 0 \\
 \frac{\gamma }{3} & 0 & 0 & 0 & \frac{1}{15} & \frac{1}{5} & \frac{1}{15} & 0 & 0 & 0 \\
 \frac{\gamma }{3} & 0 & 0 & 0 & \frac{1}{15} & \frac{1}{15} & \frac{1}{5} & 0 & 0 & 0 \\
 0 & 0 & 0 & 0 & 0 & 0 & 0 & \frac{1}{15} & 0 & 0 \\
 0 & 0 & 0 & 0 & 0 & 0 & 0 & 0 & \frac{1}{15} & 0 \\
 0 & 0 & 0 & 0 & 0 & 0 & 0 & 0 & 0 & \frac{1}{15} \\
\end{pmatrix}.
\end{equation}
We provide in \cref{eq:J_critevent} the expression for the jacobian required to compute the number density of critical events in a covariant form.
Here $x_{{ijk}}$ stands for the  derivative of the field of order $i$ w.r.t. the first direction, $j$ w.r.t. the second direction and $k$ w.r.t. the third direction divided by the 
corresponding RMS defined by \cref{eq:defRMS}.
\begin{figure*}
\begin{multline}
    \frac{J_y}{R} = \det \vb{H}_y \times (\laplacian\grad{y})\vdot \vb{H}_y^{-1}\vdot\grad{\det \vb{H}_y}=
    \\
    \bigg[y_{{}_{002}} y_{{}_{020}} y_{{}_{300}}\!-\!2 y_{{}_{002}} y_{{}_{110}} y_{{}_{210}}\!+\!y_{{}_{002}} y_{{}_{120}} y_{{}_{200}}\!+\!2 y_{{}_{011}} y_{{}_{101}} y_{{}_{210}}\!+\!2 y_{{}_{011}} y_{{}_{110}} y_{{}_{201}}\!-\!2 y_{{}_{011}} y_{{}_{111}} y_{{}_{200}}\!-\!y_{{}_{011}}^2y_{{}_{300}}\\
        -2 y_{{}_{020}} y_{{}_{101}} y_{{}_{201}}+y_{{}_{020}} y_{{}_{102}} y_{{}_{200}}+2 y_{{}_{101}} y_{{}_{110}} y_{{}_{111}}-y_{{}_{101}}^2 y_{{}_{120}}-y_{{}_{102}} y_{{}_{110}}^2
    \bigg] \\
    \times\left((y_{{}_{012}}\!+\!y_{{}_{030}}\!+\!y_{{}_{210}}) (y_{{}_{011}} y_{{}_{101}}\!-\!y_{{}_{002}} y_{{}_{110}})\!+\!(y_{{}_{002}} y_{{}_{020}}\!-\!y_{{}_{011}}^2) (y_{{}_{102}}+y_{{}_{120}}\!+\!y_{{}_{300}})\!+\!(y_{{}_{003}}\!+\!y_{{}_{021}}\!+\!y_{{}_{201}}) (y_{{}_{011}} y_{{}_{110}}\!-\!y_{{}_{020}} y_{{}_{101}})\right)+\\
    \bigg[y_{{}_{002}} y_{{}_{020}} y_{{}_{210}}\!+\!y_{{}_{002}} y_{{}_{030}} y_{{}_{200}}\!-\!2 y_{{}_{002}} y_{{}_{110}} y_{{}_{120}}\!-\!2 y_{{}_{011}} y_{{}_{021}} y_{{}_{200}}\!+\!2 y_{{}_{011}} y_{{}_{101}} y_{{}_{120}}\!+\!2 y_{{}_{011}} y_{{}_{110}} y_{{}_{111}}\!-\!y_{{}_{011}}^2 y_{{}_{210}}\\
        \!+\!y_{{}_{012}} y_{{}_{020}} y_{{}_{200}}\!-\!y_{{}_{012}} y_{{}_{110}}^2\!-\!2 y_{{}_{020}} y_{{}_{101}} y_{{}_{111}}\!+\!2 y_{{}_{021}} y_{{}_{101}} y_{{}_{110}}\!-\!y_{{}_{030}} y_{{}_{101}}^2\bigg] \\
    \times\left((y_{{}_{003}}+y_{{}_{021}}\!+\!y_{{}_{201}}) (y_{{}_{020}} y_{{}_{200}}\!-\!y_{{}_{110}}^2)\!+\!(y_{{}_{012}}\!+\!y_{{}_{030}}\!+\!y_{{}_{210}}) (y_{{}_{101}} y_{{}_{110}}-y_{{}_{011}} y_{{}_{200}})+(y_{{}_{102}}\!+\!y_{{}_{120}}\!+\!y_{{}_{300}}) (y_{{}_{011}} y_{{}_{110}}\!-\!y_{{}_{020}} y_{{}_{101}})\right)+\\
    \bigg[y_{{}_{002}} y_{{}_{020}} y_{{}_{201}}\!+\!y_{{}_{002}} y_{{}_{021}} y_{{}_{200}}-2 y_{{}_{002}} y_{{}_{110}} y_{{}_{111}}\!+\!y_{{}_{003}} y_{{}_{020}} y_{{}_{200}}\!-\!y_{{}_{003}} y_{{}_{110}}^2\!-\!2 y_{{}_{011}} y_{{}_{012}} y_{{}_{200}}+2 y_{{}_{011}} y_{{}_{101}} y_{{}_{111}}\\
        \!+\!2 y_{{}_{011}} y_{{}_{102}} y_{{}_{110}} \!-\! y_{{}_{011}}^2y_{{}_{201}}\!+\! 2 y_{{}_{012}} y_{{}_{101}} y_{{}_{110}}\!-\!2 y_{{}_{020}} y_{{}_{101}} y_{{}_{102}}\!-\!y_{{}_{021}} y_{{}_{101}}^2\bigg] \\
    \times\left((y_{{}_{102}}\!+\!y_{{}_{120}}\!+\!y_{{}_{300}}) (y_{{}_{011}} y_{{}_{101}}\!-\!y_{{}_{002}} y_{{}_{110}})\!+\!(y_{{}_{002}} y_{{}_{200}}\!-\!y_{{}_{101}}^2) (y_{{}_{012}}\!+\!y_{{}_{030}}\!+\!y_{{}_{210}})\!+\!(y_{{}_{003}}\!+\!y_{{}_{021}}\!+\!y_{{}_{201}}) (y_{{}_{101}} y_{{}_{110}}\!-\!y_{{}_{011}} y_{{}_{200}})\right).
    \label{eq:J_critevent}
\end{multline}
\end{figure*}

\subsection{Numerical implementation}\label{sec:monte_carlo}

In this section, we describe how \cref{fig:2pcf_analytical} was obtained.
We used Monte-Carlo integration to numerically evaluate \cref{eq:dNdudw}.
The statistical distribution of the aforementioned field variables is regulated by its $30\times 30$ covariance matrix $\Sigma$, which we may compute symbolically for given separation distance $r$, smoothing scales $R_{\pk}$ and $R_{\ce}$ and spectral index $n_{\rm s}$.
We aim to sample points following this distribution and evaluate the integrands of equation \eqref{eq:peak-event-char} to obtain the expectancies  $\left\langle\operatorname{Peak}(\vb*{x}) \operatorname{Event}(\vb*{y})\right\rangle$. 
Ideally, we would like to evaluate the correlation function at identical smoothing scale.
However, at equal smoothing scale and small separation, the covariance matrix becomes almost singular, resulting in unfavorable numerical artifacts plaguing the statistic of our result.
To avoid this, we instead consider slightly different smoothing scales.

Of course, we can not naively do this directly, as $x, x_i, y, y_i, \det \vb{H_y}$ will rarely take the specific values that the Dirac deltas impose, and thus the integrands will almost always evaluate to zero.
To circumvent this, we compute the distribution of the other field variables conditional to the Dirac deltas being satisfied, with two exceptions:
as we will need to integrate $\nu_\pk$ over a range to recover \cref{eq:dNdudw} anyway, we replace $\dirac(y-\nu_\pk)$ by $\heaviside(\nu_{\operatorname{peak}}^{\mathrm{max}}-y)\heaviside(y - \nu_{\operatorname{peak}}^{\mathrm{min}})$; 
furthermore, as $\det \vb{H_y}$ depends on several field variables, computing the conditional distribution is not easily feasible.
We therefore replace $\dirac(\det \vb{H_y})$ by a `thick Dirac delta', $\dirac^\varepsilon(\det \vb{H_y}) = \frac{1}{2\epsilon}\heaviside(\varepsilon-\det \vb{H_y})\heaviside(\det \vb{H_y}+\varepsilon)$, for some small $\varepsilon$. The smaller $\varepsilon$ is, the thinner the Dirac delta and the more faithful it is to the original integral. However, reducing $\varepsilon$ comes at the cost of reducing the convergence rate of the Monte-Carlo approach.

Once the conditional distribution is computed, we simply draw random points following this distribution and average the evaluation of the integrand on these points, which converges to the value of the integral.
This allows to compute $\int_{\nu_pw}^{\nu_p}\mathrm d\nu\langle\operatorname{Peak}(\vb*{x})\operatorname{Event}(\vb*{y})\rangle$ and $\langle\operatorname{Peak}(\vb*{x})\rangle$, and results in \cref{fig:2pcf_analytical}.
Practically, to obtain \cref{fig:2pcf_analytical}, we drew $2^{26}$ sample points. 
We used a smoothing ratio $R_\pk/R_\ce = 0.95$ and a thick Dirac delta of size $\varepsilon = 10^{-3}$.

\section{ Event differential distribution} \label{sec:event-distrib}

\begin{figure}
    \includegraphics[width=\columnwidth]{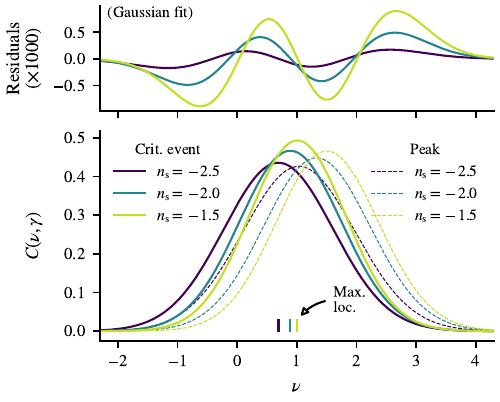}
    \caption{
        Normalized differential distribution of critical events in density rarity value for several values of spectral slope, as marked.
        The top panel shows the absolute residuals when fitting $C(\nu,\gamma)$ with a Gaussian function.
        Parameters of the Gaussian are provided in the text.
        We show the peak rarity distribution as thin dashed lines.
    }
    \label{fig:cv}
\end{figure}

The differential density of critical events with different field values $\nu$ are derived in \cite{Cadiou2020} and read 
\label{sec:diff_events}
\begin{multline}
    \label{eq:critical_event_density}
    n_\ce(\nu, R) = \frac{3R}{\tilde{R}^2R_*^3}\frac{1}{5}\left(\frac{3}{2\pi}\right)^{3/2}(1-\tilde\gamma^2)\\
    \sum_{i=5,6,9}c_{3,i}(\nu,\gamma) \exp\left(-\frac{\nu^2}{2 (1-5\gamma^2/i)}\right),
\end{multline}
where the full expressions for $c_{3,i}(\nu,\gamma)$ were given in equation~(39) of~\cite{Cadiou2020}
Factorizing the mean density of the critical events, $ \bar{n}_{\ce}$, in this expression such that
\begin{equation}
n_\ce(\nu, R) = \bar{n}_{\ce} C(\nu),   \,\, \mathrm{with}\,\,\,  \int_{-\infty}^\infty \dd{\nu} C(\nu) = 1,
\end{equation}
we introduce the normalized distribution of the events in height
\begin{equation}
    \label{eq:critical_event_Cv}
    C(\nu, \gamma) = \frac{18 \sqrt{10 \pi}}{29-6\sqrt{6}}
    \sum_{i=5,6,9}c_{3,i}(\nu,\gamma)\exp\left(-\frac{\nu^2}{2 (1-5\gamma^2/i)}\right), \notag
\end{equation}
which depends on the power spectrum (and on the smoothing scale if the spectrum is not scale invariant) only through the spectral
parameter $\gamma$. In \cref{fig:cv}, we show the behaviour of $C(\nu, \gamma)$  for several choices of $\gamma$ that correspond to the spectral
slopes of interest $n=-2.5,-2,-1.5$, see \cref{eq:defgamma}. For comparison, the peak rarity distribution is represented as dashed lines. We see that
critical events typically occur at lower $\nu$ than peaks, and are very rare for $\nu > 3$.

We note that $C(\nu, \gamma)$ for $\nu < 3$ can be approximated with an accuracy of better than \num{e-3} for $-2.5 \leq n_\mathrm{s} \leq -1.5$ by a Gaussian with parameters
\begin{align}
    \nonumber
    \tilde{C}(\nu, \gamma) &= \frac{1}{\sqrt{2\pi} \tilde{\sigma}}\exp\left(-\frac{(\nu-\tilde{\mu})^2}{2\tilde{\sigma}^2}\right)\,,\\
    \nonumber
    \tilde{\mu} &= 1.546\gamma,\,\,\,
    \tilde{\sigma} %
    = 1-0.5084\gamma^2+0.04140\gamma.
    \label{eq:gaussian_approx}
\end{align}
We show on the top panel of \cref{fig:cv} the residuals of this fit.

\section{Mean evolution of  peak rarity}\label{sec:mean_evolution}
Let us predict the evolution of peak rarity with Gaussian smoothing.
The value of the density field smoothed with a Gaussian window at position $x$
changes with window size $R$ according to the diffusion equation
\begin{equation}
\frac{\partial \delta(\vec{r},R)}{\partial R} = R \Delta \delta(\vec{r}_\pk,R)\,.
\end{equation}
If we are interested in the change of the peak value along the peak track
$\vec{r}_{\pk}(R)$, we notice that the partial derivative $\partial/\partial R$ can
be replaced by the full derivative, since at the peak
$\grad \delta = 0$, thus
\begin{equation}
\frac{\dd{\delta(\vec{r}_{\pk},R)}}{\dd{R}} = R \Delta \delta(\vec{r}_{\pk},R)\,,
\end{equation}
and for brevity, we can now drop the reference to the position of the peak
in the argument.
In terms of the rarity of the peak,
$\nu(R)=\delta(R)/\sigma_0(R)$, the evolution equation becomes
\begin{equation}
\frac{\dd{\nu(R)}}{\dd \log R} = - \nu(R) \frac{\dd\log\sigma_0}{\dd \log R} +  R^2 \Delta \nu(R)\,.
\label{eq:peakrarity}
\end{equation}
The process of changing the peak height with smoothing
is stochastic since the Laplacian
of the field is a random quantity. However, to estimate the mean change,
let us approximate the stochastic Laplacian by its conditional mean value given
the height of the peak
\begin{equation}
\Delta \nu(R) \to \left\langle \Delta \nu(R) | \nu(R), \lambda_i < 0 \right\rangle,
\label{eq:meanDelta}
\end{equation}
where the last conditional inequality, written compactly in terms
of the eigenvalues $\{\lambda_i\}_{i=1,\mathrm{D}}$ of the Hessian of the density,
ensures that we are dealing with a
maximum and not an arbitrary field point. In principle, one should also enforce
the vanishing gradient of the field at peak position, but since the gradient
of the field is uncorrelated with both the value and the second derivatives of the
field at the same point, this condition is inconsequential to our problem.

Let us first evaluate the conditional mean in \cref{eq:meanDelta} in 1D, where the
peak condition is just $\Delta \nu < 0$.
This gives us
\begin{equation}
\left\langle \Delta \nu | \nu, \Delta \nu < 0 \right\rangle
= - \frac{\nu} {R_0^2}
- \frac{1}{\zeta R_0^2}\sqrt{\frac{2}{\pi}} \frac{
\exp\left(-\frac{\nu^2\zeta^2}{2} \right)}
{1 + \mathrm{erf}\left( \frac{\nu\zeta}{\sqrt{2}}\right)}
\label{eq:meanDeltaeval},
\end{equation}
where $R_0$ is defined by \cref{eq:defR0}, while $\zeta = {\gamma}/{\sqrt{1-\gamma^2}}$ varies from zero at
$\gamma=0$, to infinity at $\gamma=1$.
The first term in \cref{eq:meanDeltaeval} is the general conditional response of the
Laplacian to the field value, while the second correction comes from restricting
the field to be at the local maximum.
For a Gaussian window function, we also have that
${\dd{\log\sigma_0}}/{\dd{R}} + {R}/{R_0^2}=0$ for any power spectrum.
We can then obtain the (mean) evolution equation for the peak rarity
\begin{equation}
\frac{\dd{\nu(R)}}{\dd\log R} =
- \frac{R^2}{\zeta R_0^2}\sqrt{\frac{2}{\pi}} \frac{
\exp\left(-\frac{\nu^2\zeta^2}{2} \right)}
{1 + \mathrm{erf}\left( \frac{\nu\zeta}{\sqrt{2}}\right)},
\label{eq:peakrarityfinal}
\end{equation}
where $R/R_0$ is a constant that depends on the power spectrum (cf \cref{sec:spectral-parameters}); it is equal to $\zeta$ for
power-law spectra.
Thus, the rarity of maxima is, on average, decreasing slowly with
smoothing scale $R$. The decrease is less pronounced for rarer peaks
and can be in first approximation neglected for $\nu \zeta > \sqrt{2}$.

For power law spectra in $D$ dimensions, $\zeta = \sqrt{(n+D)/2}$.
We obtain a criterion for insignificant rarity drift for peaks rarer than $\nu > {2}/{\sqrt{n+D}}$.
Note that this is based on generalizing the 1D result to higher dimensions.
In 2D the drift term is more cumbersome and in 3D it does not have a closed analytical form, but the structure
of the result remains the same.
\noindent
\bsp	%
\label{lastpage}
\end{document}